\documentclass[twocolumn,aps,prl,showpacs,superscriptaddress,groupedaddress]{revtex4}
\usepackage[latin9]{inputenc}
\usepackage{color}
\usepackage{amsmath}
\usepackage{amssymb}
\usepackage{graphicx}
\usepackage{esint}
\usepackage[unicode=true,pdfusetitle,
 bookmarks=true,bookmarksnumbered=false,bookmarksopen=false,
 breaklinks=true,pdfborder={0 0 1},backref=false,colorlinks=true]
 {hyperref}
\hypersetup{
 linkcolor=red,urlcolor=blue,citecolor=red,anchorcolor=blue}
\usepackage{breakurl}

\makeatletter
\@ifundefined{textcolor}{}
{%
 \definecolor{BLACK}{gray}{0}
 \definecolor{WHITE}{gray}{1}
 \definecolor{RED}{rgb}{1,0,0}
 \definecolor{GREEN}{rgb}{0,1,0}
 \definecolor{BLUE}{rgb}{0,0,1}
 \definecolor{CYAN}{cmyk}{1,0,0,0}
 \definecolor{MAGENTA}{cmyk}{0,1,0,0}
 \definecolor{YELLOW}{cmyk}{0,0,1,0}
}

\usepackage{dcolumn}\usepackage{bm}

\makeatother

\begin{document}

\title{Numerical study of localized impurity in a Bose-Einstein condensate}

\author{Javed Akram}

\email{javedakram@daad-alumni.de}

\affiliation{Institute für Theoretische Physik, Freie Universität Berlin, Arnimallee
14, 14195 Berlin, Germany}

\affiliation{Department of Physics, COMSATS, Institute of Information Technology
Islamabad, Pakistan}

\author{Axel Pelster}

\email{axel.pelster@physik.uni-kl.de}

\affiliation{Fachbereich Physik und Forschungszentrum OPTIMAS, Technische Universität
Kaiserslautern, Germany}

\date{\today}
\begin{abstract}
Motivated by recent experiments, we investigate a single $^{133}\text{Cs}$
impurity in the center of a trapped $^{87}\text{Rb}$ Bose-Einstein
condensate. Within a zero-temperature mean-field description we provide
a one-dimensional physical intuitive model which involves two coupled
differential equations for the condensate and the impurity wave function,
which we solve numerically. With this we determine within the equilibrium
phase diagram spanned by the intra- and inter-species coupling strength,
whether the impurity is localized at the trap center or expelled to
the condensate border. In the former case we find that the impurity
induces a bump or dip on the condensate for an attractive or a repulsive
Rb-Cs interaction strength, respectively. Conversely, the condensate
environment leads to an effective mass of the impurity which increases
quadratically for small interspecies interaction strength. Afterwards,
we investigate how the impurity imprint upon the condensate wave function
evolves for two quench scenarios. At first we consider the case that
the harmonic confinement is released. During the resulting time-of-flight
expansion it turns out that the impurity-induced bump in the condensate
wave function starts decaying marginally, whereas the dip decays with
a characteristic time scale which decreases with increasing repulsive
impurity-BEC interaction strength. Secondly, once the attractive or
repulsive interspecies coupling constant is switched off, we find
that white-shock waves or bi-solitons emerge which both oscillate
within the harmonic confinement with a characteristic frequency. 
\end{abstract}


\maketitle

\section{Introduction}

Recent developments in theoretical and experimental research focus
on controlling a single or few particle impurities in an ultracold
quantum gas in view of detecting and engineering strongly correlated
quantum states \cite{Gericke08,Bakr09,Sherson10,Serwane15042011}.
This research direction paths the way for a huge number of proposals
for novel applications. For example, a well-localized single atom
impurity with spin allows to study the Kondo effect \cite{Gorshkov10}.
Dressed spin-down impurities in a spin-up Fermi sea of ultracold atoms
even offer to investigate the quantum transport of spin impurity atoms
through a strongly interacting Fermi gas \cite{PhysRevLett.102.230402,PhysRevLett.103.150601}.
Furthermore, realizations of a single trapped ion impurity in a BEC
features a spatial resolution on the micrometer scale which is advantageous
in comparison with absorption imaging \cite{Zipkes10,PhysRevLett.105.133202}.
Atomtronics applications are envisioned with single atoms acting as
switches for a macroscopic system in an atomtronics circuit \cite{PhysRevLett.93.140408}.
Two impurity atoms immersed in a Bose-Einstein condensate can entangle
through phonon exchange in a quantum gas \cite{PhysRevA.71.033605},
or individual qubits can be cooled preserving internal state coherence
\cite{PhysRevA.69.022306,PhysRevLett.97.220403}. By adding impurities
one by one, experimentalists can track, in principle, the transition
from the one-body to the many-body regime, which ultimately yields
information about cluster formation \cite{Klein07}. By implementing
a single-atom within a Bose-Einstein condensate also fundamental questions
of quantum mechanics can be addressed with remarkable precision, for
instance, to which extent a single impurity can act as a local and
nondestructive probe for a strongly correlated quantum many-body state
\cite{PhysRevA.78.023610,Balewski13}. In addition, the experimental
achievement to trap a single impurity within a BEC \cite{Lercher11,Spethmann012,PhysRevLett.109.235301,Widera15}
allows for investigating polaronic physics within the realm of ultracold
quantum gases \cite{PhysRevA.84.063612,Santamore11,PhysRevA.85.023623,Grusdt2015}.

A convenient model to study the hybrid system of impurity within a
Bose-Einstein condensate (BEC) at zero-temperature relies on the mean-field
dynamics of two-coupled differential equations (DEs) for the condensate
and the impurity wave function. For the sake of simplicity, we aim
in this paper to analyze such a hybrid system in just one dimension.
This is physically justified in case that the confinement in two spatial
dimensions is much larger than the third dimension, so the 3D DEs
reduce to a truly one-dimensional (1D) or a quasi 1D model. The first
case requires transverse length scales on the order of or less than
the atomic interaction length, which is realizable near a confinement-induced
resonance \cite{PhysRevLett.81.938,PhysRevLett.85.3745,PhysRevLett.91.163201}
and allows for seminal experiments within the Tonks-Girardeau as well
as the super-Tonks-Girardeau regime \cite{Paredes04,Kinoshita20082004,Haller04092009}.
On the other hand, when the transverse confinement is larger than
the atomic interaction strength, the DEs can be reduced to an effective
quasi 1D model \cite{Kamchatnov04}. The trapping of a BEC in highly
elongated optical and magnetic traps demonstrates that a quasi-1D
BEC is experimentally realizable \cite{PhysRevA.63.031602,PhysRevLett.87.130402,Kinoshita20082004}.
Note that such a mean-field description of a quasi-1D system at zero
temperature neglects both quantum and thermal fluctuations which are
known to be enhanced within a reduced dimensionality \cite{PhysRevLett.87.130402,PhysRevLett.87.080403,PhysRevLett.87.160406,petrov04,PhysRevLett.96.130403}.
But if two-particle interaction strength and temperature are small
enough, this quasi-1D mean-field model should provide a reasonable
description.

\begin{figure}
\includegraphics[scale=0.4]{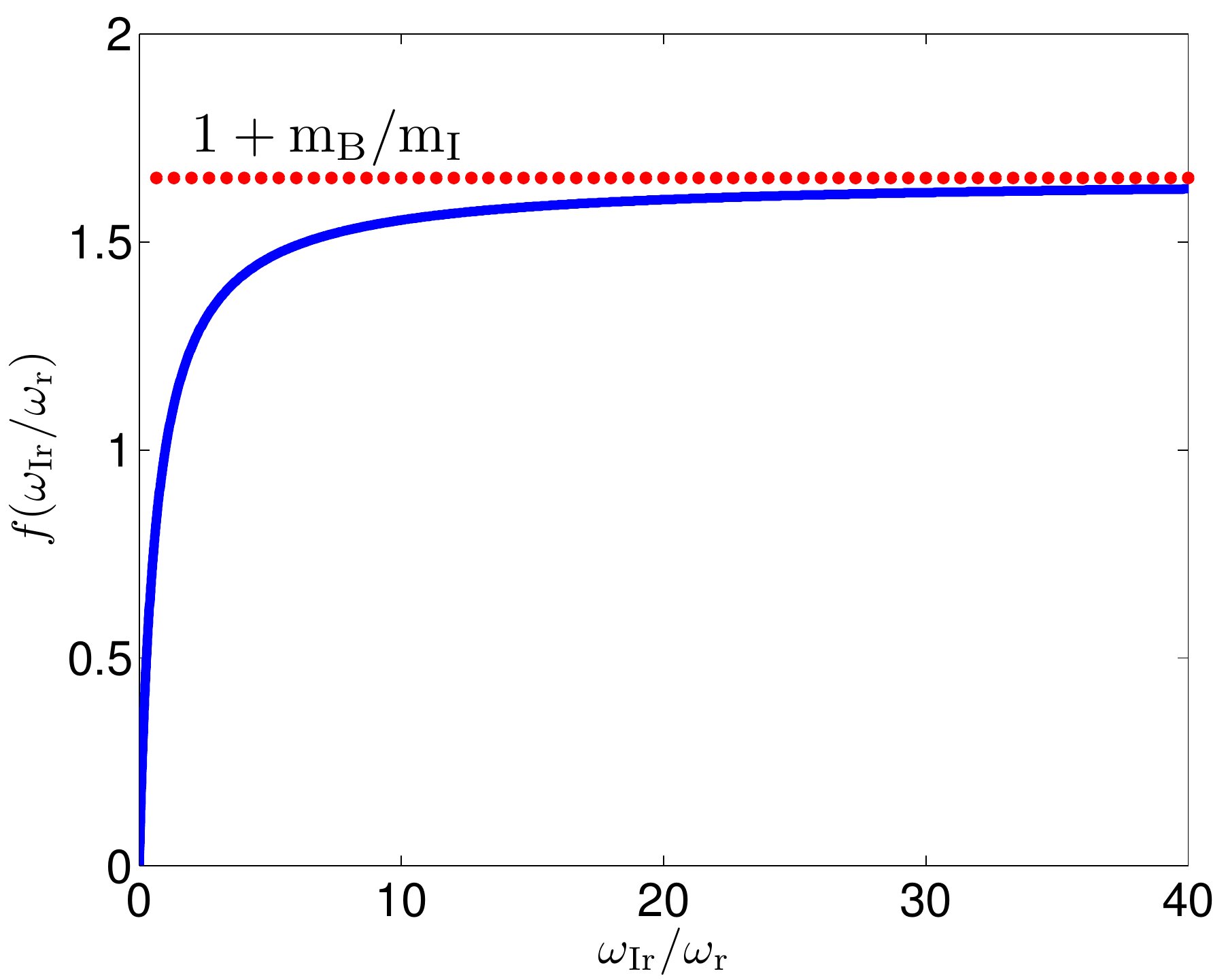}\protect\protect\protect\protect\caption{(Color online) Geometric function $f\left(\omega_{\text{Ir}}/\omega_{\text{r}}\right)$
reaches its maximum value at $1+m_{\text{B}}/m_{\text{I}}$. \label{Fig1} }
\end{figure}

Inspired by the recent experiments \cite{Lercher11,Spethmann012,PhysRevLett.109.235301,Widera15},
we propose and analyze a quasi one-dimensional model of a hybrid system,
which consists of a single $^{133}\text{Cs}$ impurity in a $^{87}\text{Rb}$
Bose-Einstein condensate. To this end, we start with defining the
underlying quasi-1D model in Sec.~\ref{C6-S2}. As a result the effective
one-dimensional interspecies coupling strength depends not only on
the three-dimensional s-wave scattering length, but also on the transversal
trap frequencies of cesium and rubidium, respectively. In the same
section, we determine the equilibrium phase diagram spanned by the
intra- and inter-species coupling strength, and specify the regions
where the impurity is localized at the trap center or expelled to
the condensate border. Afterwards in Sec.~\ref{C6-S3}, we show for
the former case that the impurity imprint upon the condensate wave
function is either a bump or a dip, depending upon whether the effective
impurity-BEC coupling strength is attractive or repulsive. Conversely,
due to the presence of the condensate, the effective mass of the impurity
turns out to depend quadratically upon a small interspecies coupling
strength. Subsequently, Sec.~\ref{C6-S4} discusses the dynamics
of the impurity imprint upon the condensate wave function for two
quench scenarios. After having released the trap, the resulting time-of-flight
expansion shows that the impurity imprint marginally decreases for
an attractive s-wave coupling but decreases for a repulsive s-wave
scattering with a characteristic time scale which decreases with increasing
the interspecies coupling strength. Furthermore, we investigate the
emergence of white-shock waves or gray/dark bi-solitons when the initial
negative or positive interspecies coupling constant is switched off.
Section~\ref{C6-S5} summarizes our findings for the proposed quasi
1D model system in view of a possible experimental realization. Finally,
in Appendix~\ref{C6-S6} we derive for the quasi one-dimensional
model the two underlying differential equations (1DDEs) for the condensate
and the impurity wave function from a three-dimensional setting.

\begin{figure*}[t]
\begin{centering}
\includegraphics[width=16cm,height=5.0cm]{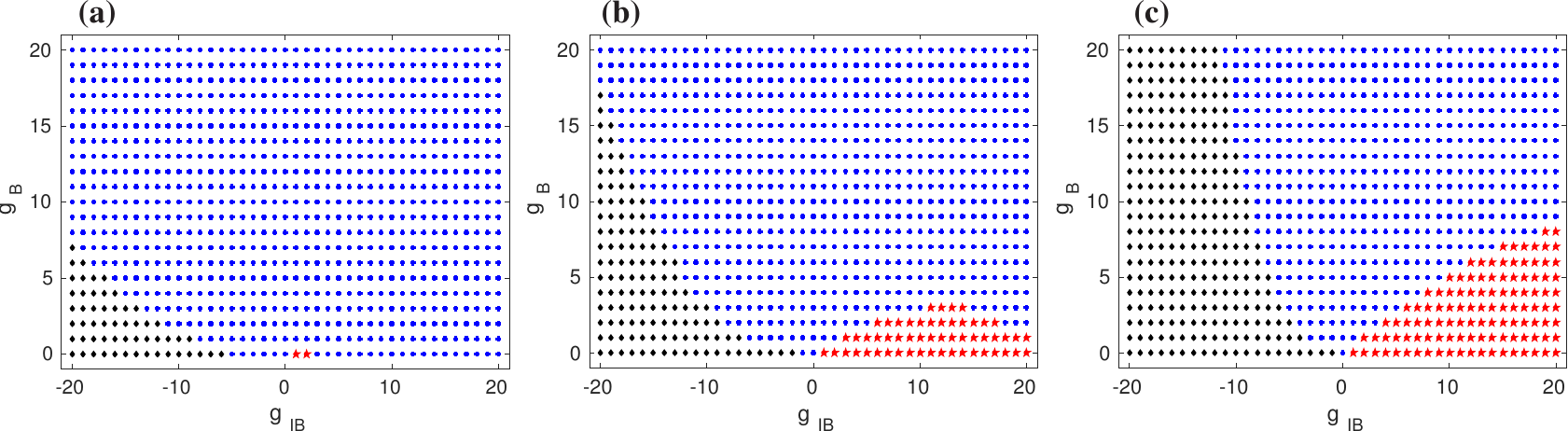} 
\par\end{centering}

\protect\protect\protect\protect\protect\caption{(Color online) Equilibrium phase diagram spanned by $g_{\text{B}}$
and $g_{\text{IB}}$ for  (a) $N_{\text{B}}=20$  (b) $N_{\text{B}}=200$
and  (c) $N_{\text{B}}=800$. Impurity is localized at trap center (blue)
or expelled to the condensate border (red) together with unstable
region (black) in dimensionless units. \label{Fig2}}
\end{figure*}

\begin{figure*}
\includegraphics[width=14cm,height=9.5cm]{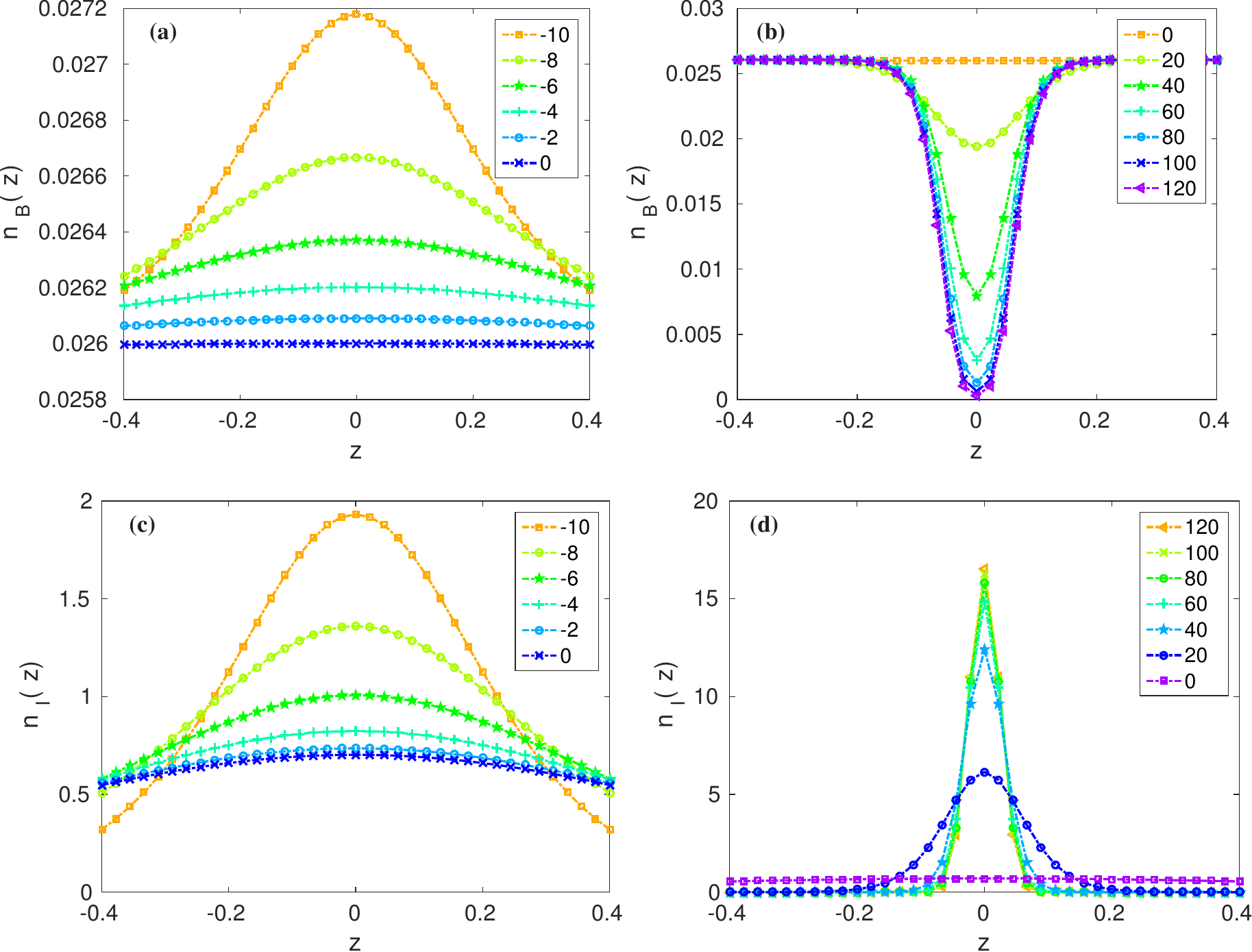} \protect\protect\protect\protect\protect\caption{(Color online) Numerical density profile of BEC (a-b) and impurity
(c-d) for two-particle Rb-Rb coupling constant value $G_{\text{B}}=16000$
and for interspecies coupling constants $g_{\text{IB}}$ which increases
from top to bottom according to the inlets. For increasing negative
values of $g_{\text{IB}}$ the impurity-induced bump (a) in the condensate
wave function decreases, whereas for positive values the corresponding
dip (b) increases  in dimensionless units. \label{Fig3}}
\end{figure*}

\begin{figure*}[th]
\includegraphics[width=16cm,height=6.0cm]{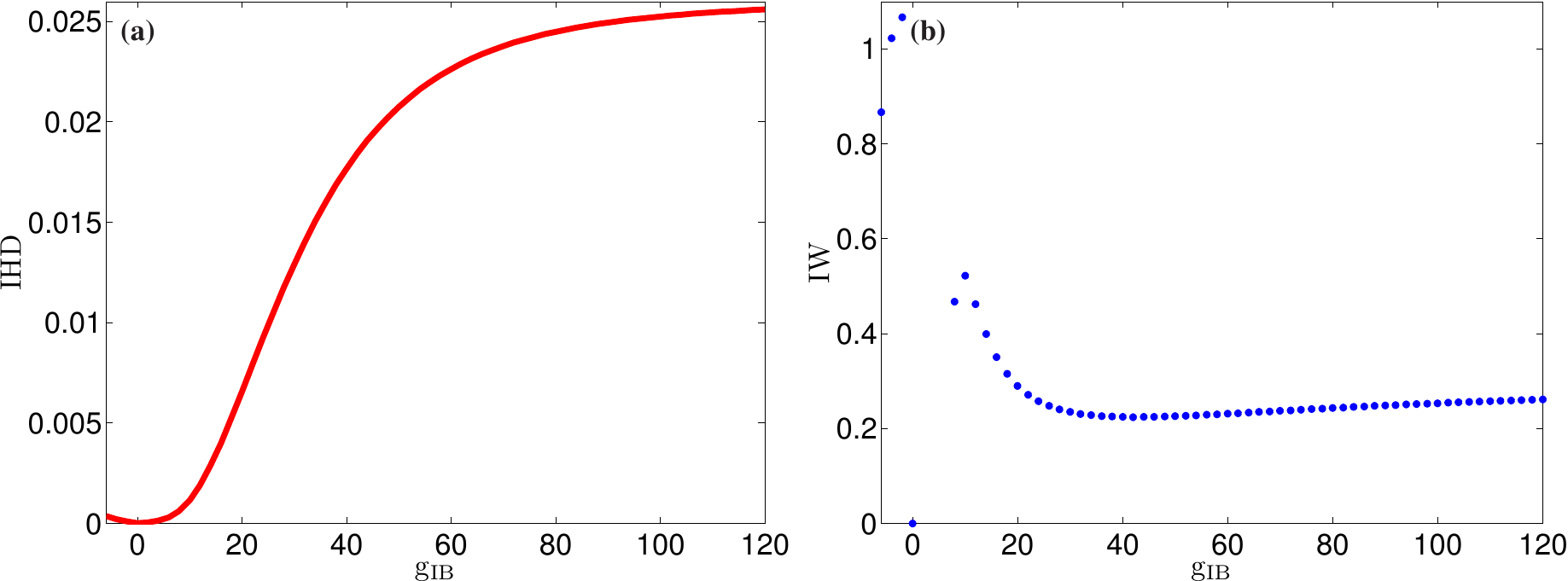}\protect\protect\protect\protect\protect\protect\caption{(Color online)  (a) Height/depth and  (b) width of impurity bump/dip according
to Eqs.~(\ref{eq20})--(\ref{eq21}) versus impurity-BEC coupling
constant $g_{\text{IB}}$ for the BEC coupling constant $G_{\text{B}}=16000$
calculated numerically by solving 1DDEs (\ref{eq11}) and (\ref{eq:11-1})  in dimensionless units.
\label{Fig4}}
\end{figure*}

\begin{figure}
\includegraphics[width=8.5cm,height=6.0cm]{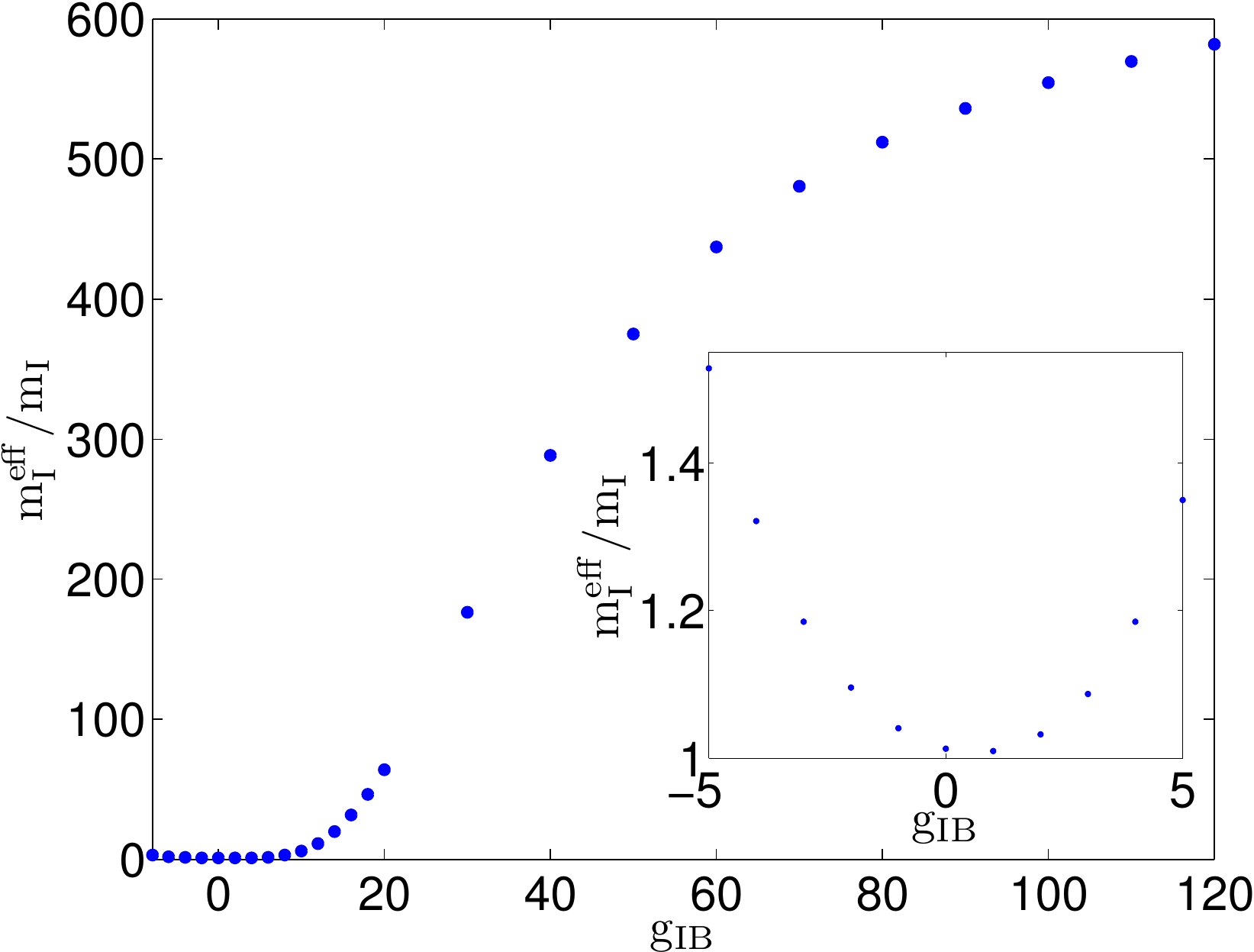} \protect\protect\protect\protect\protect\protect\caption{(Color online) Effective mass of $^{133}\text{Cs}$ impurity versus
the impurity-BEC coupling strength $g_{\text{IB}}$. Inlet shows that
effective mass increases quadratically for small impurity-BEC coupling
strength $g_{\text{IB}}$  in dimensionless units.\label{Fig5} }
\end{figure}

\section{Quasi 1D model}

\label{C6-S2}We start with considering a BEC and an impurity confined
in a harmonic trap. In case of a strong transversal confinement one
obtains an effective quasi one-dimensional model which is described
by two-coupled one-dimensional differential equations for the underlying
condensate and impurity wave function $\psi_{\text{B}}\left(z,t\right)$
and $\psi_{\text{I}}(z,t)$, respectively. Appendix~\ref{C6-S6}
outlines how the one-dimensional Lagrangian density (\ref{eq4}) is
obtained from an original three-dimensional setting by integrating
out the two transversal degrees of freedom. The Euler-Lagrangian equations
(\ref{eq9}) then reduce to the two coupled one-dimensional differential
equations (1DDEs): 
\begin{align}
i\hbar\frac{\partial}{\partial t}\psi_{\text{B}}\left(z,t\right) & =\left\{ -\frac{\hbar^{2}}{2m_{\text{B}}}\frac{\partial^{2}}{\partial z^{2}}+\frac{m_{\text{B}}\omega_{\text{z}}^{2}}{2}z^{2}+G_{\text{IB}}\parallel\psi_{\text{I}}(z,t)\parallel^{2}\right.\nonumber \\
 & +G_{\text{B}}\parallel\psi_{\text{B}}\left(z,t\right)\parallel^{2}\biggr\}\psi_{\text{B}}\left(z,t\right),\label{eq10}\\
i\hbar\frac{\partial}{\partial t}\psi_{\text{I}}(z,t) & =\left\{ -\frac{\hbar^{2}}{2m_{\text{I}}}\frac{\partial^{2}}{\partial z^{2}}+\frac{m_{\text{I}}\omega_{\text{Iz}}^{2}}{2}+G_{\text{BI}}\parallel\psi_{\text{B}}\left(z,t\right)\parallel^{2}\right\} \nonumber \\
 & \times\psi_{\text{I}}(z,t).\label{eq:10-1}
\end{align}
On the right-hand side of Eqs. (\ref{eq10}) and (\ref{eq:10-1})
the first term represents the kinetic energy of the BEC(impurity)
atoms with mass $m_{\text{B}}$($m_{\text{I}}$), the second term
describes the potential energy term, the third term stands for the
impurity-BEC coupling with the respective strengths $G_{\text{IB}}=g_{\text{IB}}$,
$G_{\text{BI}}=N_{\text{B}}g_{\text{IB}}$, and the last term in (\ref{eq10})
represents the Rb-Rb two-particle interaction with strength $G_{\text{B}}=N_{\text{B}}g_{\text{B}}$.
In Appendix~\ref{C6-S6} it is determined how $g_{\text{B}}$ and
$g_{\text{IB}}$ depend on the s-wave scattering lengths $a_{\text{B}}$
and $a_{\text{IB}}$. For intraspecies coupling one gets 
\begin{align}
g_{\text{B}} & =2a_{\text{B}}\hbar\omega_{\text{r}},\label{eq:10-2}
\end{align}
whereas for interspecies coupling one obtains 
\begin{align}
g_{\text{IB}} & =2a_{\text{IB}}\hbar\omega_{\text{r}}f\left(\frac{\omega_{\text{Ir}}}{\omega_{\text{r}}}\right).\label{eq:10-3}
\end{align}
Here the geometric function 
\begin{align}
f\left(\frac{\omega_{\text{Ir}}}{\omega_{\text{r}}}\right) & =\frac{1+\left(m_{\text{B}}/m_{\text{I}}\right)}{1+\left(m_{\text{B}}\omega_{\text{r}}\right)/\left(m_{\text{I}}\omega_{\text{Ir}}\right)},\label{eq:10-4}
\end{align}
depends on the ratio of the trap frequencies as depicted in Fig.~\ref{Fig1}.
Thus, $f\left(\omega_{\text{Ir}}/\omega_{\text{r}}\right)$ is monotonously
increasing, equals to one for the present case $\omega_{\text{Ir}}=\omega_{\text{r}}$,
and reaches its maximum value at $1+m_{\text{B}}/m_{\text{I}}$ for
the frequency ratio of about $\omega_{\text{Ir}}/\omega_{\text{r}}\geq20$.
In order to vary the impurity-BEC coupling strength there are, in
principle, two possibilities according to Eq. (\ref{eq:10-4}): either
the ratio of the radial trap frequencies is tuned as shown in Fig.~\ref{Fig1},
or the s-wave scattering length $a_{\text{IB}}$ is modified with
the use of a Feshbach resonance \cite{Lercher11,PhysRevA.79.042718,PhysRevA.85.032506}.
In order to make Eqs. (\ref{eq10}) and (\ref{eq:10-1}) dimensionless
we introduce the dimensionless time as $\tilde{t}=\omega_{\text{z}}t$,
the dimensionless coordinate $\tilde{z}=z/l_{\text{z}}$, and the
dimensionless wave function $\tilde{\psi}=\psi\sqrt{l_{\text{z}}}$
with the oscillator length $l_{\text{z}}=\sqrt{\hbar/(m_{\text{B}}\omega_{\text{z}})}$.
With this Eqs. (\ref{eq10}) and (\ref{eq:10-1}) can be rewritten
in the form 

\begin{figure*}
\includegraphics[width=16cm,height=6.5cm]{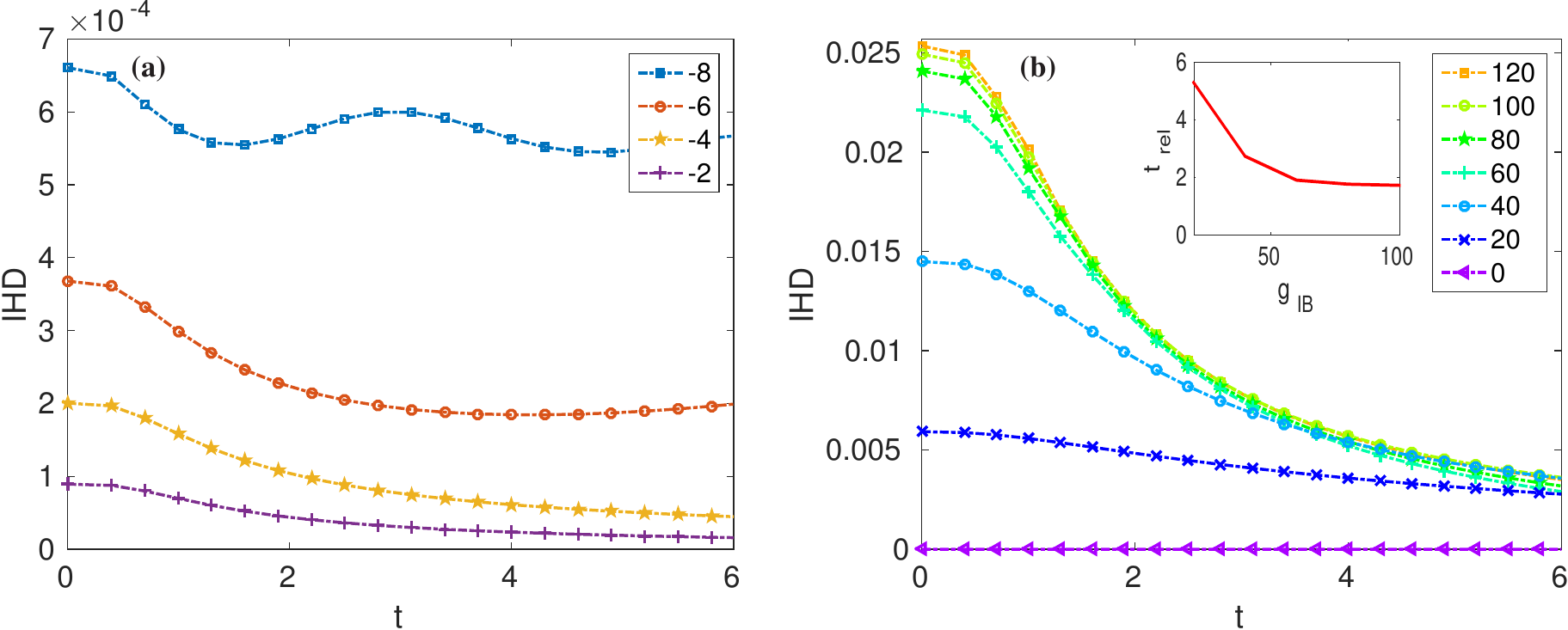} \protect\protect\protect\protect\protect\protect\caption{(Color online) Impurity imprint height/depth after having released
the trap versus time for  (a) increasing negative and  (b) decreasing
positive values of impurity-BEC coupling constant $g_{\text{IB}}$
from top to bottom. Inlet: relaxation time $t_{\text{rel}}$ decreases
with increasing $g_{\text{IB}}$  in dimensionless units. \label{Fig06}}
\end{figure*}

\begin{align}
i\frac{\partial}{\partial\tilde{t}}\tilde{\psi}_{\text{B}}\left(\tilde{z},\tilde{t}\right) & =\left\{ -\frac{1}{2}\frac{\partial^{2}}{\partial\tilde{z}^{2}}+\frac{\tilde{z}^{2}}{2}+\tilde{G}_{\text{IB}}\parallel\tilde{\psi}_{\text{I}}(\tilde{z},\tilde{t})\parallel^{2}\right.\nonumber \\
 & +\tilde{G}_{\text{B}}\parallel\tilde{\psi}_{\text{B}}\left(\tilde{z},\tilde{t}\right)\parallel^{2}\biggr\}\tilde{\psi}_{\text{B}}\left(\tilde{z},\tilde{t}\right),\label{eq11}\\
i\frac{\partial}{\partial\tilde{t}}\tilde{\psi}_{\text{I}}(\tilde{z},\tilde{t}) & =\left\{ -\frac{\tilde{\alpha}^{2}}{2}\frac{\partial^{2}}{\partial\tilde{z}^{2}}+\frac{\tilde{z}^{2}}{2\tilde{\alpha}^{2}}+\tilde{G}_{\text{BI}}\parallel\tilde{\psi}_{\text{B}}(\tilde{z},\tilde{t})\parallel^{2}\right\} \nonumber \\
 & \times\tilde{\psi}_{\text{I}}(\tilde{z},\tilde{t}),\label{eq:11-1}
\end{align}
where we have $\tilde{G}_{\text{B}}=N_{\text{B}}\tilde{g}_{\text{B}}$,
$\tilde{G}_{\text{IB}}=\tilde{g}_{\text{IB}}$ and $\tilde{G}_{\text{BI}}=N_{\text{B}}\tilde{g}_{\text{IB}}$
with $\tilde{g}_{\text{B}}=g_{\text{B}}/\left(\hbar\omega_{\text{z}}l_{\text{z}}\right)$,
and $\tilde{g}_{\text{IB}}=g_{\text{IB}}/\left(\hbar\omega_{\text{z}}l_{\text{z}}\right)$.
Here $\tilde{\alpha}=l_{\text{Iz}}/l_{\text{z}}$ defines the ratio
of the two oscillator lengths. Thus, we can summarize that Eq. (\ref{eq11})
is nothing but a standard Gross-Piteavskii equation with an additional
potential stemming from the impurity, whereas Eq. (\ref{eq:11-1})
is a typical Schrödinger wave equation with a potential originating
from the BEC. From here on, we will drop the tildes for simplicity.

Typically, a mixture of two species can occur in two different states,
either it is miscible, i.e. both species overlap, or it is immiscible,
i.e. the two species do not overlap \cite{PhysRevLett.80.1130,Sartori13}. In
our case the equilibrium phase diagram is spanned by the coupling
strengths $g_{\text{B}}$ and $g_{\text{IB}}$ and contains a region,
where the impurity is localized at the center, and another one, where
the impurity is expelled to be localized at the border of the condensate.
Note that a similar equilibrium phase diagram was studied for the
homogeneous case with attractive interspecies s-wave scattering lengths
in Ref. \cite{PhysRevA.73.043608}. In order to investigate the physical
regions of interest for our proposed model, we solve the two coupled
1DDEs (\ref{eq11}) and (\ref{eq:11-1}) in imaginary time numerically
by using the split-operator method \cite{Vudragovic12,Kumar15,Loncar15,Sataric16},
which yields the equilibrium phase diagram Fig.~\ref{Fig2}. The
blue region shows where the impurity is localized at the center of
the BEC, the red region depicts that the impurity is displaced from
the center to the border of the condensate, and finally, the black
region represents the unstable region where impurity and condensate
do not coexist.

In the rest of the paper, we are interested into the localization
of the impurity at the center of the BEC, therefore, from now on,
we only focus on the blue region in the equilibrium phase diagram
of Fig.~\ref{Fig2}. In particular, we consider that the BEC consists
of $N_{\text{B}}=800$ $^{87}\text{Rb}$ atoms, for the dimensionless
intraspecies couplings constant we assume $G_{\text{B}}=16000$, and
we let the ratio of the two oscillator lengths $\alpha=l_{\text{Iz}}/l_{\text{z}}$
to have the value 0.808.

\section{Impurity Imprint Upon Stationary Condensate Wave Function}

\label{C6-S3}In order to determine the impurity imprint on the condensate
wave function in equilibrium we solve the two coupled 1DDEs numerically
in imaginary time. In this way we find that the impurity leads to
a bump/hole in the BEC density at the trap center for negative/positive
values of $g_{\text{IB}}$ as shown in Fig.~\ref{Fig3}. For increasing
the attractive/repulsive interspecies coupling strength the bump/dip
upon the condensate decreases/increases. For the repulsive interspecies
coupling strength, the impurity drills a dip in the BEC density which
gets deeper and deeper until no more BEC atoms remain in the trap
center and, finally, the BEC fully fragments into two parts as shown
in Fig.~\ref{Fig3}(b) at the characteristic value $g_{\text{IBc}}=110$.
The width/height of the impurity wave function decreases/increases
for increasing interspecies coupling constant $\left|g_{\text{IB}}\right|$,
respectively, as shown in Fig.~\ref{Fig3}(c-d).

\begin{figure}
\includegraphics[width=8.5cm,height=6.0cm]{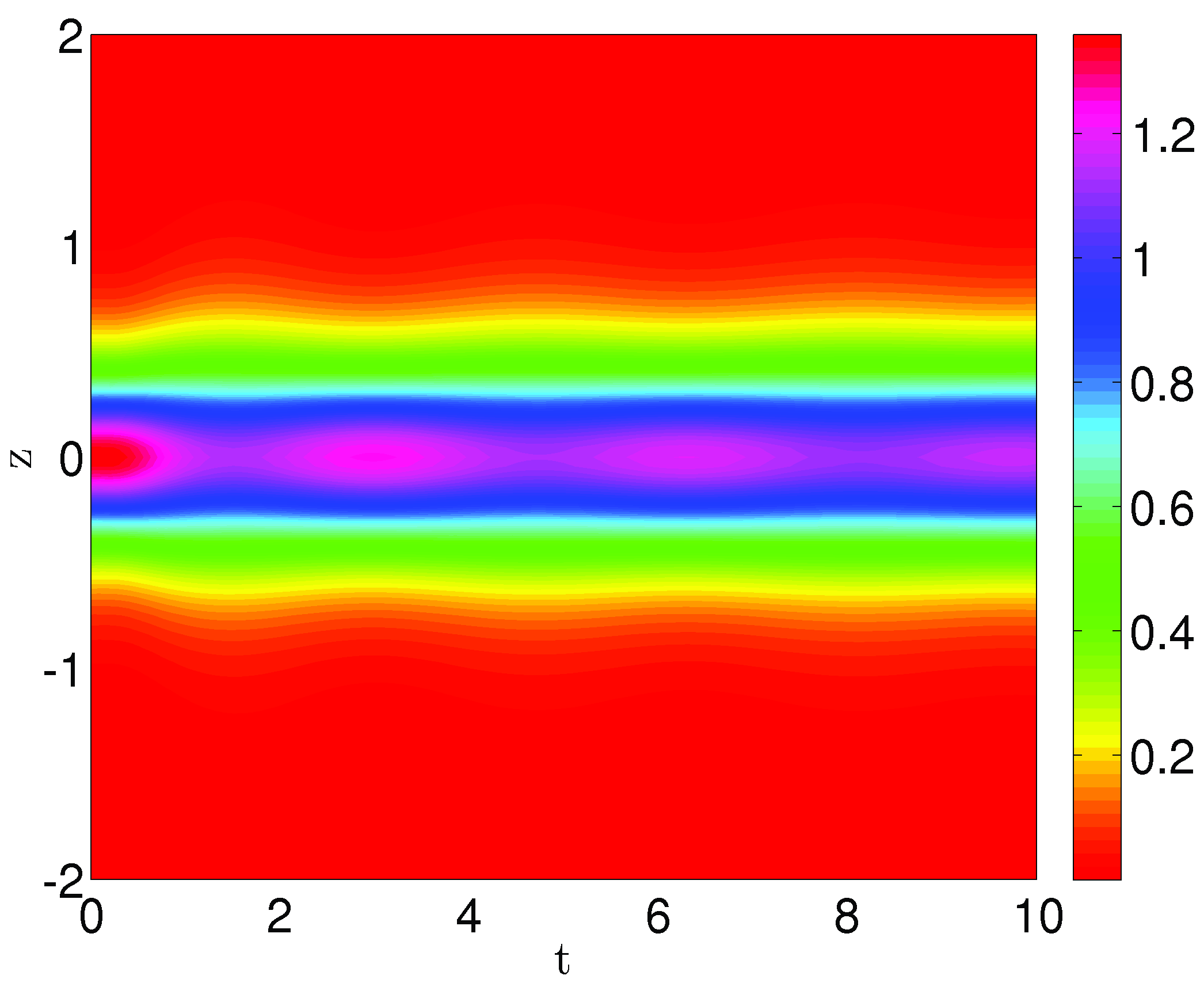} \protect\protect\protect\protect\protect\protect\caption{(Color online) Dynamics of impurity density represented in color scale
after having switched off the harmonic trap for initial $g_{\text{IB}}$=-8
for the BEC coupling strength $G_{\text{B}}=16000$ in dimensionless units. \label{Fig7}}
\end{figure}

\begin{figure}
\includegraphics[width=8.5cm,height=6.0cm]{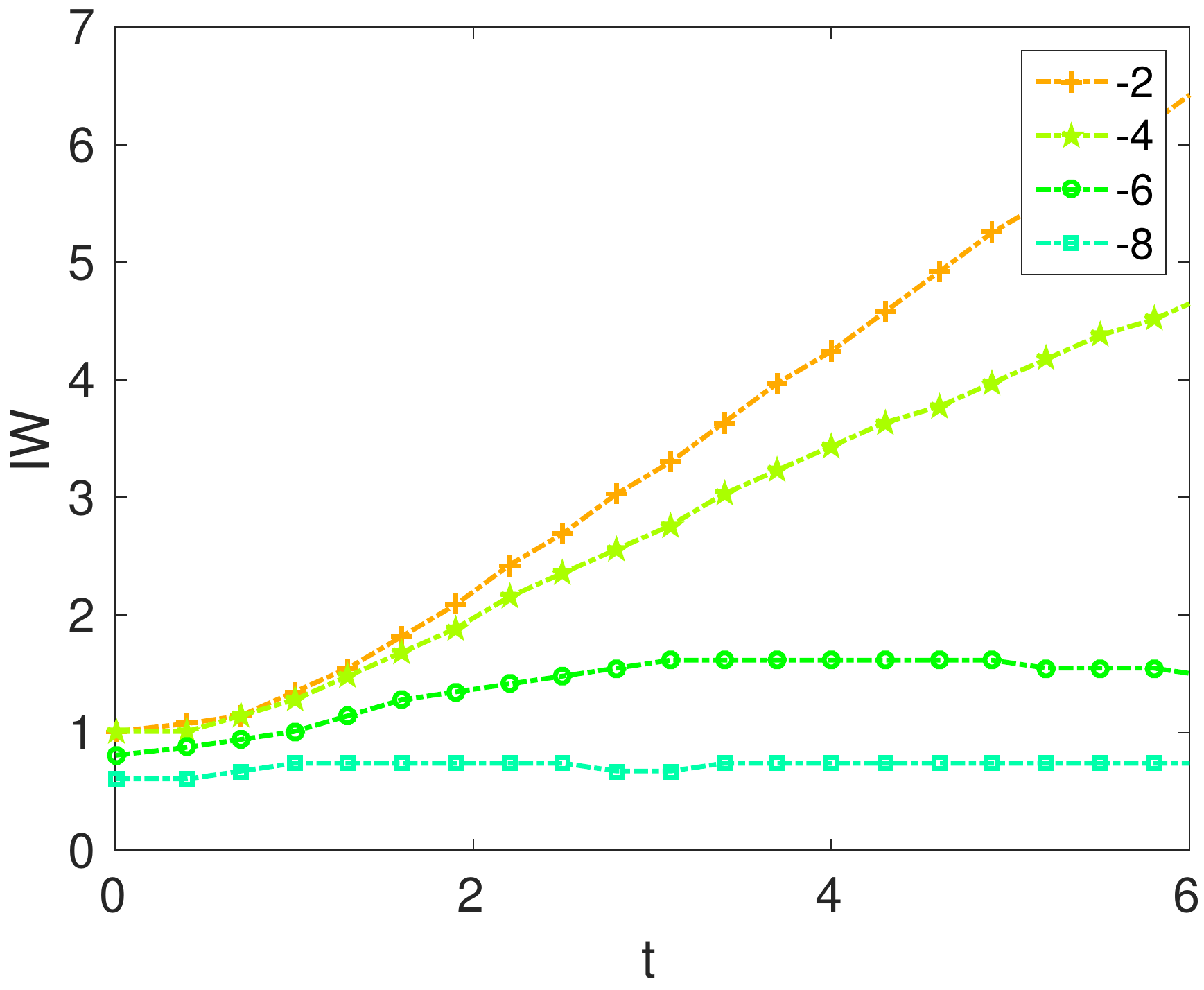} \protect\protect\protect\protect\protect\protect\caption{(Color online) Impurity imprint width after having released the trap
versus time for decreasing positive values of impurity-BEC coupling
constant $g_{\text{IB}}$ from top to bottom  in dimensionless units. \label{Fig8}}
\end{figure}

In view of a more detailed comparison, we describe the impurity imprint
upon the condensate wave function $\psi_{\text{B}}(z)$ by the following
two quantities. The first one is the impurity height/depth 
\begin{eqnarray}
 & \begin{array}{r@{}l@{\,}l}
\text{IHD}=\left\{ \begin{array}{r@{}l@{\,}l}
\parallel\psi_{\text{B}}\left(0\right)\parallel_{g_{\text{IB}}}^{2}-\parallel\psi_{\text{B}}\left(0\right)\parallel_{g_{\text{IB}}=0}^{2}\qquad g_{\text{IB}}\leq0\\
\,\\
\text{Max}\left(\parallel\psi_{\text{B}}\left(z\right)\parallel_{g_{\text{IB}}}^{2}\right)-\parallel\psi_{\text{B}}\left(0\right)\parallel_{g_{\text{IB}}}^{2}\qquad g_{\text{IB}}\geq0
\end{array}\right.\end{array}\label{eq20}
\end{eqnarray}
and the second one is the impurity width $\text{IW}$, which we define
as follows. For $g_{\text{IB}}\leq0$ we use the full width half maximum
\begin{eqnarray}
\parallel\psi\left(\text{IW}/2\right)\parallel_{g_{\text{IB}}}^{2}=\frac{\parallel\psi_{\text{B}}\left(0\right)\parallel_{g_{\text{IB}}}^{2}+\parallel\psi_{\text{B}}\left(0\right)\parallel_{g_{\text{IB}}=0}^{2}}{2}\,\: g_{\text{IB}}\leq0\,,
\end{eqnarray}
whereas for $g_{\text{IB}}\geq0$ we define the equivalent width \cite{carroll2007}:
\begin{eqnarray}
\text{IW}=\frac{2I_{\text{0}}z_{\text{Max}}-\intop_{-z_{\text{Max}}}^{z_{\text{Max}}}\parallel\psi_{\text{B}}\left(z\right)\parallel_{g_{\text{IB}}}^{2}dz}{I_{\text{0}}-\parallel\psi_{\text{B}}\left(0\right)\parallel_{g_{\text{IB}}}^{2}}\quad g_{\text{IB}}\geq0\,,\label{eq21}
\end{eqnarray}
where we have $I_{\text{0}}=\text{Max}\left(\parallel\psi_{\text{B}}\left(z\right)\parallel_{g_{\text{IB}}}^{2}\right)$.
In Fig.~\ref{Fig4}(a) we plot the IHD for $g_{\text{IB}}>-10$,
while $g_{\text{IB}}<-10$ is not a valid region for $g_{\text{B}}=20$
according to Fig.~\ref{Fig2}(c). From Figure~\ref{Fig4}(a) we
read off that for $g_{\text{IB}}=0$, i.e.~when there is no impurity
present, the impurity height/depth vanishes. The IHD quadratically
increases for the repulsive interspecies coupling strength $0<g_{\text{IB}}<60$
and partially fragments the BEC until it reaches its marginally saturated
value $\text{IHDc}\approx0.025$ for the characteristic interspecies
coupling strength $g_{\text{IBc}}=110$. In the case of $g_{\text{IB}}>g_{\text{IBc}}$,
the impurity fully fragments the BEC into two parts as shown in Fig.~\ref{Fig3}(b).
The impurity imprint width increases abruptly just before/after $g_{\text{IB}}=0$
for attractive/repulsive interspecies coupling strength, respectively,
as shown in Fig.~\ref{Fig4}(b). For an increasing repulsive impurity-BEC
coupling strength the impurity width then decreases until it reaches
the interspecies coupling strength $g_{\text{IB}}=30$, later on it
marginally increases until the characteristic interspecies coupling
strength $g_{\text{IBc}}=110$, where we have $\text{IWc}\approx0.23$.

The effective mass of the impurity is defined as $m_{\text{I}}^{\text{eff}}=\hbar/\left(l_{\text{Iz}}^{2}\omega_{\text{z}}\right)$,
where the impurity oscillator length $l_{\text{Iz}}=\sqrt{2}\sigma$
follows from the standard deviation $\sigma=\sqrt{<z^{2}>-<z>^{2}}$,
with $<\bullet>=\int\bullet\left|\psi_{\text{I}}(z)\right|^{2}dz$
denoting the expectation value. Figure \ref{Fig5} shows the ratio
of the effective mass of the $^{133}\text{Cs}$ impurity with respect
to the bare mass $m_{\text{I}}$, which increases quadratically for
interspecies coupling strength $-5<g_{\text{IB}}<5$ as shown in the
inlet of Fig.~\ref{Fig5}, and becomes marginally saturated for interspecies
coupling strength $g_{\text{IB}}>g_{\text{IBc}}$. Note that our results
for the effective mass of the impurity are restricted to the mean-field
regime. In order to go beyond and include the impact of quantum fluctuations,
one would need to investigate polaron physics \cite{PhysRevB.80.184504,PhysRevA.84.063612,Santamore11,Grusdt2015}.

\section{Impurity Imprint Upon Condensate Dynamics}

\label{C6-S4}In an experiment, any imprint of the impurity upon the
condensate wave function could only be detected dynamically. Thus,
it is of high interest to study theoretically whether the impurity
imprint, which we have found and analyzed for the stationary case
in the previous section, remains present also during the dynamical
evolution of the condensate wave function. To this end, we explore
two quench scenarios numerically in more detail. The first one is
the standard time-of-flight (TOF) expansion after having switched
off the external trap when the interspecies interaction is still present.
In the second case we consider the inverted situation that the impurity-BEC
interaction is suddenly switched off within a remaining harmonic confinement,
which turns out to give rise to the emergence of wave packets or bi-solitons
depending on whether the initial interspecies interaction strength
is attractive or repulsive.

\subsection{Time-of-Flight Expansion}

Time-of-flight (TOF) absorption pictures represent an important diagnostic
tool to analyze dilute quantum gases since the field's inception.
By suddenly turning off the magnetic trap, the atom cloud expands
non-ballistically with a dynamics which is determined by both the
momentum distribution of the atoms at the instance, when the confining
potential is switched off, and by inter-atomic interactions \cite{PhysRevLett.77.416,Inouye99}.
We have investigated the time-of-flight expansion dynamics of the
BEC with impurity by solving numerically the two coupled 1DDEs (\ref{eq11}),
(\ref{eq:11-1}) and analyzing the resulting evolution of both the
condensate and the impurity wave function. It turns out that, despite
the continuous broadening of the condensate density, its impurity
imprint remains qualitatively preserved both for attractive and repulsive
interspecies interaction strengths, respectively. Therefore, we focus
a more quantitative discussion upon the dynamics of the corresponding
impurity height/depth and width.

For an attractive Rb-Cs coupling strength, it turns out that the impurity
imprint even remains approximately constant in the time-of-flight,
as is shown explicitly in Fig.~\ref{Fig06} (a) for the IHD, which
marginally decreases for $t\gg0$. As shown in Fig.~\ref{Fig06}
(a) for smaller attractive interspecies coupling strength, we observe
the crumbling breathing of the impurity upon IHD as discussed recently
for the Bose-Hubbard model \cite{PhysRevLett.110.015302}. For the
attractive interspecies coupling strength $g_{\text{IB}}=-8$ the
dynamics of the impurity density is shown in Fig.~\ref{Fig7}, which
clearly reflects the crumbling breathing of the impurity at the center
of the BEC. In case of the IW, we find that the IW starts increasing
marginally for smaller values of attractive interspecies coupling
strength and increases linearly for larger attractive interspecies
coupling strength $g_{\text{IB}}$ as shown in Fig.~\ref{Fig8}.

\begin{figure}
\includegraphics[width=8.5cm,height=10cm]{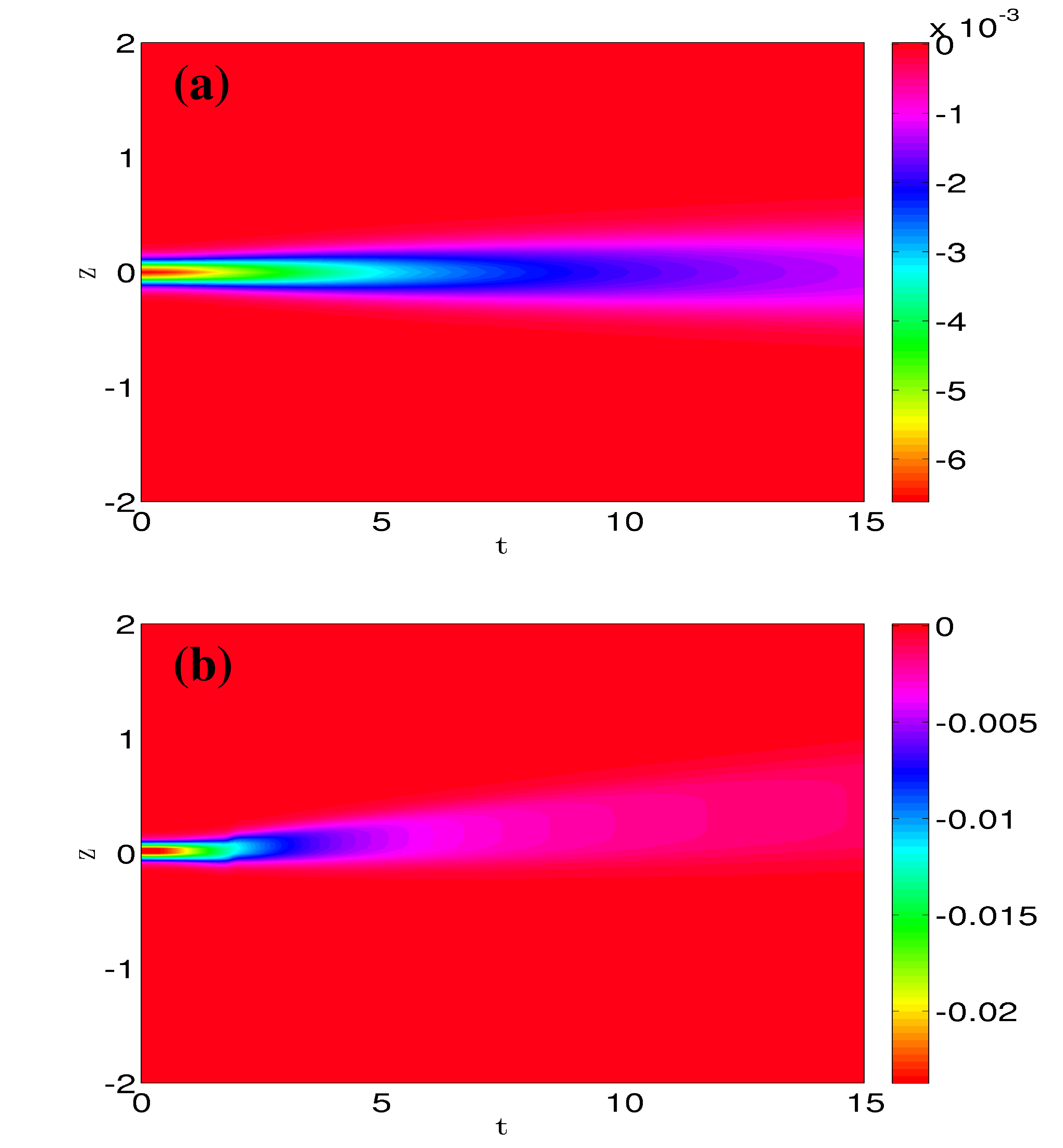} \protect\protect\protect\protect\protect\protect\caption{(Color online) Coherent matter-wave time-of-flight evolution of the
depleted density $\left\Vert \psi_{\text{B}}\left(z,t\right)\right\Vert _{\text{\text{DD}}}^{2}=\left\Vert \psi_{\text{B}}\left(z,t\right)\right\Vert _{\text{\ensuremath{g_{\text{IB}}}}}^{2}-\left\Vert \psi_{\text{B}}\left(z,t\right)\right\Vert _{\text{\ensuremath{g_{\text{IB}}=0}}}^{2}$
represented in color scale after having switched off the trap for
the BEC coupling constant $G_{\text{B}}=16000$:  (a) $g_{\text{IB}}=20$
and  (b) $g_{\text{IB}}=80$ in dimensionless units. \label{Fig9}}
\end{figure}

In case of a repulsive interspecies interaction, the IHD decays with
a characteristic time scale as shown in Fig.~\ref{Fig06}(b).  Defining
that relaxation time $t_{\textrm{rel}}$ according to $\textrm{IHD}(t_{\textrm{rel}})=\textrm{IHD}(0)/2$,
the inlet reveals that the impurity imprint depth relaxes with a shorter
time scale for increasing repulsive impurity-BEC coupling strength.
This physical picture is confirmed by the time-of-flight evolution
of the depleted density depicted in Fig.~\ref{Fig9}. At the beginning
of TOF the impurity-imprint remains at first constant, then the imprint
width expands and the imprint height decays faster for a larger 
interspecies coupling strength. In Fig.~\ref{Fig9} we plotted the
time-of-flight of the depleted density of the BEC for two cases. 
For the repulsive interspecies coupling strength $g_{\text{IB}}=20$
we observe that the impurity imprint decays marginally from its equilibrium
value as shown in Fig.~\ref{Fig06}(b) and the impurity remains localized
at the trap center as shown in Fig.~\ref{Fig9}(a). On the other
hand, for the larger value of the repulsive interspecies coupling
strength $g_{\text{IB}}=80$, the impurity imprint decays from its
equilibrium value as shown in Fig.~\ref{Fig06}(b) and at the same
time the impurity is expelled from the center of the BEC as shown
in Fig.~\ref{Fig9}(b). With this we conclude that for a small enough
repulsive interspecies coupling strengths the impurity survives in
the center of the BEC for larger times.

\begin{figure}
\includegraphics[width=8.5cm,height=6.0cm]{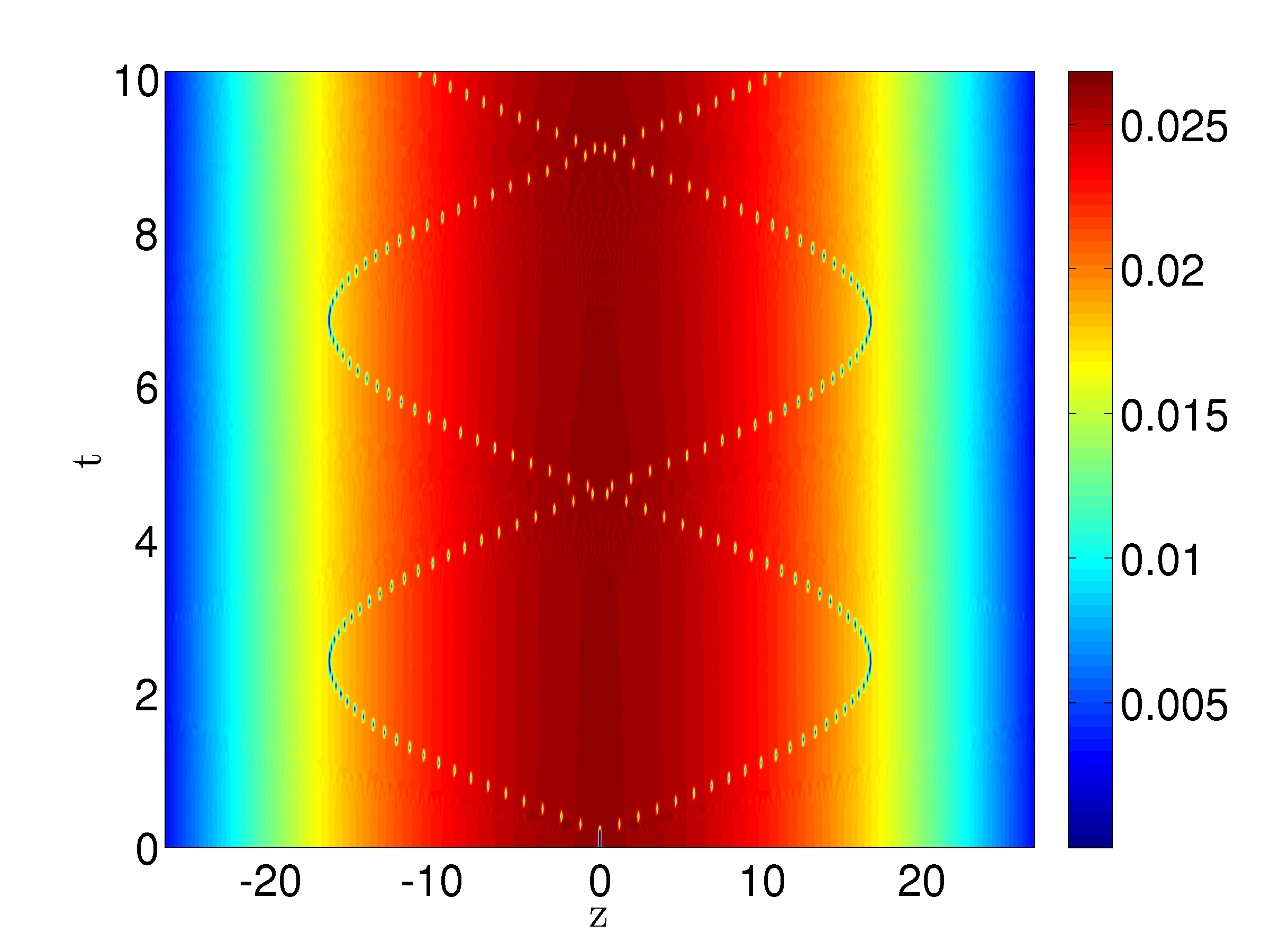} \protect\protect\protect\protect\protect\protect\caption{(Color online) Density profile of BEC represented in color scale after
having switched off the impurity-BEC coupling constant for initial
$g_{\text{IB}}$=120 for the BEC coupling constant $G_{\text{B}}=16000$ in dimensionless units.
\label{Fig10}}
\end{figure}

\subsection{Wave Packets Versus Solitons}

Due to their quantum coherence, BECs exhibit rich and complex dynamic
patterns, which range from the celebrated matter-wave interference
of two colliding condensates \cite{Andrews97} over Faraday waves
\cite{PhysRevLett.98.095301,PhysRevA.85.023613} to the particle-like
excitations of solitons \cite{Reinhardt97,Kivshar199881,Scott98,PhysRevLett.84.2298,PhysRevA.63.043613,Becker08,Shomroni09,Javed15,Javed15-2}.
For our proposed quasi 1D model of a BEC with an impurity we investigated
the dynamics of the condensate wave function which emerges after having
switched off the interspecies coupling strength. Both for an initial
attractive and repulsive interspecies coupling strength $g_{\text{IB}}$
we observe that two excitations of the condensate are created at the
impurity position, which travel in opposite direction with the same
center-of-mass speed, are reflected at the trap boundaries and then
collide at the impurity position as shown exemplarily in Fig.~\ref{Fig10}
for the initial $g_{\text{IB}}$=120 and $G_{\text{B}}=16000$. These
excitations qualitatively retain their shape despite the collision
at the impurity position. All these findings are not yet conclusive
to decide whether these excitations represent wave packets in the
absence of dispersion or solitons. Therefore, we investigate their
dynamics in more detail, by determining their center-of-mass motion
via \cite{Javed15} 
\begin{equation}
\bar{z}_{{\rm {L,R}}}\left(t\right)=\frac{\int_{-\infty,0}^{0,\infty}z\left(\parallel\psi_{\text{B}}\left(z,t\right)\parallel_{g_{\text{IB}}}^{2}-\parallel\psi_{\text{B}}\left(z,t\right)\parallel_{g_{\text{IB}}=0}^{2}\right)dz}{\int_{-\infty,0}^{0,\infty}\left(\parallel\psi_{\text{B}}\left(z,t\right)\parallel_{g_{\text{IB}}}^{2}-\parallel\psi_{\text{B}}\left(z,t\right)\parallel_{g_{\text{IB}}=0}^{2}\right)dz}\,,\label{eq25}
\end{equation}

\begin{figure*}
\includegraphics[width=17cm,height=5.5cm]{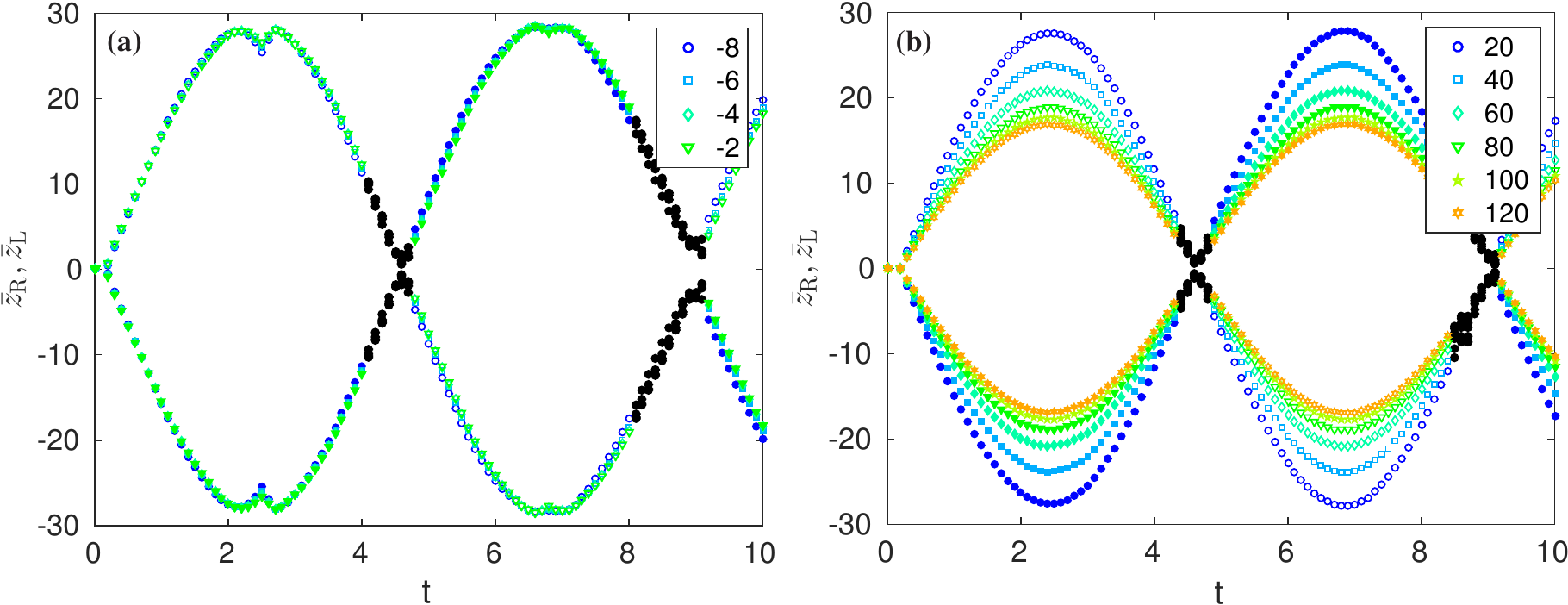} \protect\protect\protect\protect\protect\protect\caption{(Color online) Center of mass positions of excitations $\bar{z}_{{\rm L}}$
(filled circles) and $\bar{z}_{{\rm R}}$ (empty circles) according
to Eq. (\ref{eq25}) versus time after having switched off  (a) negative
decreasing and  (b) positive increasing values of $g_{\text{IB}}$ from
top to bottom in dimensionless units. Black filled circles represent the region of colliding
excitations, where mean positions are not perfectly detectable. \label{Fig11}}
\end{figure*}

which are plotted in Fig.~\ref{Fig11}. Note that the mean positions
$\bar{z}_{{\rm {L}}}$ and $\bar{z}_{{\rm {R}}}$ of the excitations
are uncertain in the region where they collide. Nevertheless Fig.~\ref{Fig11}
demonstrates that the excitations oscillate with the frequency $\omega=2\pi\times35.7~{\rm Hz}$
irrespective of sign and size of $g_{\text{IB}}$. As we have assumed
the trap frequency $\omega_{{\rm z}}=2\pi\times50~{\rm Hz}$, we obtain
the ratio $\omega/\omega_{{\rm z}}\approx0.714$, which is quite close
to $\omega/\omega_{{\rm z}}=1/\sqrt{2}\approx0.707$. Despite these
similarities of the cases of an initial attractive and repulsive interspecies
coupling constant $g_{\text{IB}}$, we observe one significant difference.
Whereas the oscillation amplitudes of the excitations do not depend
on the value of the initial $g_{\text{IB}}<0$ according to Fig.~\ref{Fig11}
 (a), we find decreasing oscillation amplitudes of the excitations with
increasing the initial $g_{\text{IB}}>0$ in Fig.~\ref{Fig11} b).
Such an amplitude dependence on the initial condition is characteristic
for gray/dark solitons according to Ref.~\cite{PhysRevLett.84.2298}.
This particle-like interpretation of the excitations agrees with the
other theoretical prediction of Ref.~\cite{PhysRevLett.84.2298}
that gray/dark solitons oscillate in a harmonic confinement with the
frequency $\omega/\omega_{{\rm z}}=1/\sqrt{2}$, which was already
confirmed in the Hamburg experiment of Ref.~\cite{Becker08} and
is also seen in Fig.~\ref{Fig11}. Conversely, for an initial attractive
interspecies coupling constant the excitations can not be identified
with bright solitons as the dynamics is governed by a GPE with a repulsive
two-particle interaction. Here the excitations have to be interpreted
as wave packets which move without any dispersion, thus, for $g_{\text{IB}}<0$
the excitations propagate like sound waves in the BEC \cite{PhysRevLett.79.553,Javed15,Javed15-2}.
Thus, we conclude that switching off the interspecies coupling constant
leads for $g_{\text{IB}}<0$ and $g_{\text{IB}}>0$ to physically
different situations. For an initial attractive RbCs coupling constant
we generate wave packets which correspond to white-shock waves \cite{PhysRevA.73.043601},
whereas for the corresponding repulsive case bi-solitons emerge \cite{Strecker02,PhysRevLett.89.200404},
which are due to the collision of the two partially/fully fragmented
parts of the condensate.  When the shock wave approaches the
trap boundary, it penetrates into the trap potential, which leads
to a change of its shape, later it recovers its shape. Therefore we
see a small kink in the center of mass position amplitude at time
$t\thicksim2.5$ as shown in Fig.~\ref{Fig11}(a). On the
other hand the solitary wave is characterized by preserving its shape
so it is reflected back from the potential boundary without changing
its shape. Thus we do not see any kink in Fig.~\ref{Fig11}(b). Note
that it can be shown in our proposed system that gray bi-solitons
are generated for a partially fragmented BEC, i.e. the impurity-BEC
coupling strength $g_{\text{IB}}<g_{\text{IBc}}$. On the other hand
the dark bi-solitons turn out to be only generated for $g_{\text{IB}}>g_{\text{IBc}}$,
where the BEC is fully fragmented into two parts at equilibrium. Recently,
we have observed that bi-solitons trains are generated in the traditional
harmonic trap with an additional dimple trap \cite{Javed15}. We emphasize
that beside many practical application of the impurity-BEC system,
someone can also generate solitons by considering an impurity as a
drilling appliance to fragment the BEC, which allows to study solitons
physics in the condensate.

\section{Summary and Conclusion }

\label{C6-S5}In the present work we studied within a quasi 1D model
numerically how a single impurity in the center of a trapped BEC affects
the condensate wave function. At first, we investigated the equilibrium
properties of that hybrid system by numerically solving the underlying
two coupled 1DDEs (\ref{eq11}) and (\ref{eq:11-1}) with the imaginary-time
propagation method. For an increasing attractive/repulsive Rb-Cs interaction
strength it turns out that the impurity imprint bump/dip decreases/increases
quadratically and reaches its marginally saturated value after $g_{\text{IBc}}=110$.
Later we found that the impurity imprint width increases abruptly
for increasing the attractive/repulsive Rb-Cs interaction strength,
but for the repulsive case it reaches a marginally saturated value
for $g_{\text{IB}}>g_{\text{IBc}}$. Beyond the characteristic value
$g_{\text{IBc}}$, the BEC fragments into two parts and, if $g_{\text{IB}}$
is increased beyond $g_{\text{IBc}}$, the impurity yields a condensate
wave function whose impurity width increases further, although the
impurity height/depth remains constant. Afterwards, we investigated
the impurity imprint upon the condensate dynamics for two quench scenarios.

At first, we considered the release of the harmonic confinement, which
leads to a time-of-flight expansion and found that the impurity imprint
upon the condensate decays slowly for small valves of the attractive/repulsive
interspecies coupling strength. This result suggests that it might
be experimentally easier to observe the impurity imprint for small
attractive/repulsive coupling constant $g_{\text{IB}}$. We also observed
the decaying breathing of the impurity at the center of the condensate
for small attractive Rb-Cs coupling strength. Additionally, we found
for stronger repulsive interspecies coupling strength that the $^{87}\text{Rb}$
atoms repel the single $^{133}\text{Cs}$ impurity from the center.
In an experiment one has to take into account that inelastic collisions
lead to two- and three-body losses of the condensate atoms \cite{PhysRevLett.85.1795,Gonzaez08}.
As such inelastic collisions are enhanced for a higher BEC density,
they play a vital role for an attractive interspecies coupling, when
the condensate density has a bump at the impurity position, but are
negligible in the repulsive case with the dip in the wave function.

In addition, we analyzed the condensate dynamics after having switched
off the interspecies coupling strength. This case turned out to be
an interesting laboratory in order to study the physical similarities
and differences of bright shock waves and gray/dark bi-solitons, which
emerge for an initial negative and positive interspecies coupling
constant $g_{\text{IB}}$, respectively. We consider the astonishing
observation, that the oscillation frequencies of both the shock waves
and the soliton coincide, to be an artifact of the harmonic confinement.
Additionally, we also found that the generation of gray/dark bi-solitons
is a generic phenomenon on collisions of partially/fully fragmented
BEC, respectively, which is strongly depending upon the equilibrium
values of the impurity wave function height and width.

\subsection*{Acknowledgment}

We thank James Anglin, Antun Balaž, Herwig Ott, Ednilson Santos, and
Artur Widera for insightful comments. Furthermore, we gratefully acknowledge
financial support from the German Academic Exchange Service (DAAD).
This work was also supported in part by the German-Brazilian DAAD-CAPES
program under the project name ``Dynamics of Bose-Einstein Condensates
Induced by Modulation of System Parameters\textquotedblright{} and
by the German Research Foundation (DFG) via the Collaborative Research
Center SFB/TR49 ``Condensed Matter Systems with Variable Many-Body
Interactions\textquotedblright{}.

\appendix

\section{\;}

\label{C6-S6}We start with the fact that the underlying equations
for describing an impurity immersed in a BEC can be formulated in
terms of the Hamilton principle of least action with the action functional
$\mathcal{A}_{\text{3D}}=\int dt\int\mathcal{L_{\text{3D}}}\; d^{3}r$,
where the Lagrangian density reads for three dimensions 
\begin{align}
 & \mathcal{L}_{\text{3D}}=\sum_{j=\text{B,I}}N_{\text{j}}\biggl\{\frac{i\hbar}{2}\left[\psi_{\text{j}}^{\star}\left(\mathbf{r},t\right)\frac{\partial\psi_{\text{j}}\left(\mathbf{r},t\right)}{\partial t}-\psi_{\text{j}}\left(\mathbf{r},t\right)\frac{\partial\psi_{\text{j}}^{\star}\left(\mathbf{r},t\right)}{\partial t}\right]\nonumber \\
 & +\frac{\hbar^{2}}{2m_{\text{j}}}\psi_{\text{j}}^{\star}\left(\mathbf{r},t\right)\bigtriangleup\psi_{\text{j}}\left(\mathbf{r},t\right)-V_{\text{j}}(\mathbf{r})\psi_{\text{j}}^{\star}\left(\mathbf{r},t\right)\psi_{\text{j}}\left(\mathbf{r},t\right)-\frac{N_{\text{j}}g_{\text{j}}^{\text{3D}}}{2}\nonumber \\
 & \times\parallel\psi_{\text{j}}\left(\mathbf{r},t\right)\parallel^{4}\biggr\}-N_{\text{B}}N_{\text{I}}g_{\text{IB}}^{\text{3D}}\parallel\psi_{\text{I}}\left(\mathbf{r},t\right)\parallel^{2}\parallel\psi_{\text{B}}\left(\mathbf{r},t\right)\parallel^{2}.\label{eq1}
\end{align}
Here $\psi_{\text{B}}\left(\mathbf{r},t\right)$ and $\psi_{\text{I}}\left(\mathbf{r},t\right)$
describe the BEC and the impurity wave-function with $\mathbf{r}=\left(x,\; y,\; z\right)$,
$V_{\text{B}}(\mathbf{r})=m_{\text{B}}\omega_{\text{z}}^{2}z^{2}/2+m_{\text{B}}\omega_{\text{r}}^{2}\left(x^{2}+y^{2}\right)/2$
and $V_{\text{I}}\left(\mathbf{r}\right)=m_{\text{I}}\omega_{\text{Iz}}^{2}z^{2}/2+m_{\text{I}}\omega_{\text{Ir}}^{2}\left(x^{2}+y^{2}\right)/2$
denote the three-dimensional harmonic potential for the bosons and
the $^{133}\text{Cs}$ impurity. The three-dimensional $^{87}\text{Rb}$
coupling constant reads $g_{\text{B}}^{\text{3D}}=4\pi\hbar^{2}a_{\text{B}}/m_{\text{B}}$,
where the s-wave scattering length is $a_{\text{B}}=94.7~{\rm {a}_{0}}$
with the Bohr radius ${\rm {a}_{0}}$ and $m_{\text{B}}$ stands for
the mass of the $^{87}\text{Rb}$ atom, while the three-dimensional
$^{133}\text{Cs}$ coupling constant reads $g_{\text{I}}^{\text{3D}}=0$,
because their is only one single $^{133}\text{Cs}$ impurity atom
present into the system, i.e. $N_{\text{I}}=1$. The three-dimensional
effective Rb-Cs coupling constant is $g_{\text{IB}}^{\text{3D}}=2\pi\hbar^{2}a_{\text{IB}}/m_{\text{IB}}$,
where $m_{\text{IB}}=m_{\text{I}}m_{\text{B}}/\left(m_{\text{I}}+m_{\text{B}}\right)$
is the reduced mass of two species, $m_{\text{I}}$ is the mass of
the $^{133}\text{Cs}$ atom, and $a_{\text{IB}}=650~{\rm {a}_{0}}$
represents the effective Rb-Cs s-wave scattering length \cite{Lercher11}.
We assume an effective one-dimensional setting with $\omega_{\text{z}}\ll\omega_{\text{r}}$,
so we decompose the BEC wave-function $\psi_{\text{B}}(\mathbf{r},t)=\psi_{\text{B}}(z,t)\phi_{\text{B}}(\textbf{ r}_{\perp},t)$
with $\textbf{ r}_{\perp}=\left(x,\; y\right)$ and %
\begin{eqnarray}
\phi_{\text{B}}(\textbf{ r}_{\perp},t) & = & \frac{e^{-\frac{x^{2}+y^{2}}{2l_{\text{r}}^{2}}}}{\sqrt{\pi}l_{\text{r}}}e^{-i\omega_{\text{r}}t}\,.\label{eq2}
\end{eqnarray}
Furthermore, we assume that the single impurity in the center of the
BEC is trapped by a harmonic potential with $\omega_{\text{Iz}}\ll\omega_{\text{Ir}}$.
Thus, we perform a similar decomposition of the impurity wave function
$\psi_{\text{I}}(\mathbf{r},t)=\psi_{\text{I}}(z,t)\phi_{\text{I}}(\textbf{ r}_{\perp},t)$
with 
\begin{eqnarray}
\phi_{\text{I}}(\textbf{ r}_{\perp},t) & = & \frac{e^{-\frac{x^{2}+y^{2}}{2l_{\text{rI}}^{2}}}}{\sqrt{\pi}l_{\text{rI}}}e^{-i\omega_{\text{rI}}t}\,.\label{eq3}
\end{eqnarray}
Here $l_{\text{r}}=\sqrt{\hbar/(m_{\text{B}}\omega_{\text{r}})}$
and $l_{\text{Ir}}=\sqrt{\hbar/(m_{\text{I}}\omega_{\text{Ir}})}$
denote the oscillator lengths in radial direction for BEC and impurity.
For the experimentally realistic trap frequencies $\omega_{\text{r}}=\omega_{\text{Ir}}=2\pi\times0.179~\text{kHz}\ll\omega_{\text{z}}=\omega_{\text{Iz}}=2\pi\times0.050~\text{kHz}$
\cite{Spethmann012} these radial oscillator lengths amount to the
values $l_{\text{r}}=15190.8\,{a}_{0}$ and $l_{\text{Ir}}=12279.0\,{a}_{0}$
for BEC and impurity, respectively.  In order to distinguish
between the weakly interacting quasi-1D and the strongly interacting
Tonks-Girardeau regime, Petrov et al. \cite{PhysRevLett.85.3745}
introduced a dimensionless quantity which involves both the longitudinal
and the transversal trap size as well as the scattering length: 
\begin{equation}
\alpha=2a_{\text{B}}\frac{l_{\text{z}}}{l_{\text{r}}^{2}}\,.\label{eq31}
\end{equation}
By using above mentioned experimental parameters, we get the dimensionless
quantity $\alpha=0.023,$ so we are far in the weakly interacting
regime, where the Gross-Pitaevskii mean-field theory is applicable, see also
Fig.~5 of Ref.~\cite{petrov04} and Refs.~\cite{PhysRevA.54.656,Lincoln00}. Therefore, we can follow
Ref.~\cite{Kamchatnov04} and integrate out the two transversal dimensions
of our three-dimensional Lagrangian according to $\mathcal{L}_{\text{1D}}=\int_{-\infty}^{\infty}\int_{-\infty}^{\infty}\mathcal{L}_{\text{3D}}\; dxdy$.
After some algebra, the resulting quasi one dimensional Lagrangian
reads

\begin{align}
 & \mathcal{L}_{\text{1D}}=\sum_{j=\text{B,I}}N_{\text{j}}\biggl\{\frac{i\hbar}{2}\left[\psi_{\text{j}}^{\star}\left(z,t\right)\frac{\partial\psi_{\text{j}}\left(z,t\right)}{\partial t}-\psi_{\text{j}}\left(z,t\right)\frac{\partial\psi_{\text{j}}^{\star}\left(z,t\right)}{\partial t}\right]\nonumber \\
 & +\frac{\hbar^{2}}{2m_{\text{j}}}\psi_{\text{j}}^{\star}\left(z,t\right)\bigtriangleup\psi_{\text{j}}\left(z,t\right)-V_{\text{j}}(z)\psi_{\text{j}}^{\star}\left(z,t\right)\psi_{\text{j}}\left(z,t\right)-\frac{N_{\text{j}}g_{\text{j}}}{2}\nonumber \\
 & \parallel\psi_{\text{j}}\left(z,t\right)\parallel^{4}\biggr\}-N_{\text{B}}N_{\text{I}}g_{\text{IB}}\parallel\psi_{\text{I}}\left(z,t\right)\parallel^{2}\parallel\psi_{\text{B}}\left(z,t\right)\parallel^{2},\!\!\!\!\label{eq4}
\end{align}
where $V_{\text{B,I}}\left(z\right)=m_{\text{B,I}}\omega_{\text{z,Iz}}^{2}z^{2}/2$
represents the one-dimensional harmonic potential for the BEC and
for the impurity, the one-dimensional intraspecies coupling strength
$g_{\text{B}}$ is given by (\ref{eq:10-2}) and the interspecies
coupling strength $g_{\text{IB}}$ turns out to be (\ref{eq:10-3}).
The two coupled time dependent differential equations follow from
the action $\mathcal{A}_{\text{1D}}=\int_{-\infty}^{\infty}\int_{-\infty}^{\infty}\mathcal{L}_{\text{1D}}dzdt$
and by using the Euler-Lagrangian equation 
\begin{align}
 & \frac{\partial\mathcal{L}_{\text{1D}}}{\partial\psi_{\text{j}}^{\star}\left(z,t\right)}-\frac{\partial}{\partial z}\frac{\partial\mathcal{L}_{\text{1D}}}{\partial\frac{\partial\psi_{\text{j}}^{\star}\left(z,t\right)}{\partial z}}-\frac{\partial}{\partial t}\frac{\partial\mathcal{L}_{\text{1D}}}{\partial\frac{\partial\psi_{\text{j}}^{\star}\left(z,t\right)}{\partial t}}=0.\label{eq9}
\end{align}
Inserting the one-dimensional Lagrangian density (\ref{eq4}), after
some algebra we obtain the two coupled 1DDEs (\ref{eq10}) and (\ref{eq:10-1}).

\bibliographystyle{apsrev4-1}
\bibliography{TwoCoupledRbCs.bib}

\end{document}